\documentclass[12pt]{article}
\usepackage{color}
\usepackage{latexsym}
\usepackage{amsmath}
\usepackage{amsfonts}
\usepackage{amssymb}
\usepackage{indentfirst} 
\usepackage{tikz}

\title{New superintegrable  models on spaces of constant curvature}
\author{Cezary Gonera\thanks{e-mail: cgonera@uni.lodz.pl}, \quad Joanna Gonera\thanks{e-mail: joanna.gonera@uni.lodz.pl}\\
\small Faculty of Physics and Applied Informatics, \\
\small University of \L\'od\'z,\\
\small Pomorska 149/153, 90-236 {\L}\'od\'z, Poland
}
\date{}
\begin{document}
\maketitle 
\begin{abstract}

It is known that the fairly (most?) general class of 2D superintegrable systems defined on 2D spaces of constant curvature and separating in (geodesic) polar coordinates is specified by two types of radial potentials (oscillator or (generalized) Kepler ones) and by corresponding families of angular potentials. Unlike the radial potentials the angular ones are given implicitly (up to a function) by ,in general, transcendental  equation.\\
In the present paper new two-parameter families of angular potentials are constructed in terms of elementary functions. It is shown that for an appropriate choice of parameters the family corresponding to the oscillator/Kepler  type radial potential reduces to Poschl-Teller potential. This allows to consider Hamiltonian systems defined by this family as a generalization of Tremblay-Turbiner-Winternitz (TTW) or Post-Winternitz (PW) models both on plane as well as on curved spaces of constant curvature.


\end{abstract}
\section{Introduction}
\par A Hamiltonian system defined on two-dimensional configuration space admitting three globally defined functionally independent constants of motion is called superintegrable. On classical level, for confining potentials the existence of such three integrals of motion makes all bounded trajectories closed\cite{l1}. Probably, the best known examples of such models are isotropic harmonic oscillator and Kepler-Coulomb systems.\\
The above rough definition of two-dimensional superintegrability extends to systems with n degrees of freedom and can be, if needed, made mathematically sound (see Refs.\cite{l2}, the reviews \cite{l3}, as well as references therein for relevant definitions and general setting).\\ 
Various properties and aspects of 2d-superintegrable systems both classical and quantum, in flat and curved spaces have been studied in many papers \cite{l4} - \cite{l31}.\\
In particular, several recent articles devoted to superintegrable systems in Euclidean space $E_2$ concentrated on derivation and 
classification of Hamiltonians that allow separation in polar coordinates and admit an additional polynomial in momenta integral of motion 
\cite{l32} - \cite{l37}. The main advantage of the assumption that all constants of motion are polynomials in momenta of some finite degree $N$ is that the results are valid practically both in classical and quantum regime. However, the price to be payed is that potentially interesting superintegrable models not allowing the polynomial integral constant of motion remain beyond the scheme. Especially, as derived in Ref.\cite{l6} in the case of Euclidean plane, rederived (by using different method) and generalized to 2d spaces of constant curvature in Ref.\cite{l29} superintegrable Hamiltonians defined on such spaces and separating in polar (geodesic) coordinates (see eq.(\ref{r3}) below) can be described as follows. There are two types of allowed radial potentials (isotropic harmonic oscillator and generalized Kepler type, see eq.(\ref{r15}) and (\ref{r16}) below, respectively)and two corresponding families of angular potentials. The period of the angular potential in the given family is determined by the type of corresponding radial potential. Actually, all allowed angular potentials corresponding to the radial potential of a given type exhibit the same dependence $T(L)$ of the period on a separation constant L (playing a role of the total "energy" in a bounded angular motion)- they are isoperiodic. It is known \cite{l38} that given the period of motion as a function of energy allows to reconstruct (under mild assumptions) the potential up to a practically arbitrary function. Therefore, it is rather not surprising that the polynomially superintegrable models form a subset of the superintegrable systems described above. In particular, as observed in \cite{l37} the Hamiltonian with generalized Kepler potential on Euclidean plane does not admit the third polynomial integral of motion.\\
The present paper is devoted to the further study of superinegrable Hamiltonians defined on $2d$ spaces of constant curvature and separating in polar (geodesic) coordinates. We derive (in terms of elementary functions) two-parameter families of angular potentials (corresponding to oscillator and generalized Kepler type radial potentials, respectively) generating superintegrable dynamics. This is done in the framework adopted in the Refs.\cite{l6}, \cite{l29} without assuming from the very beginning any specific form of additional constant of motion. In particular, the family corresponding to the radial potentials of oscillator/Kepler type can be seen as 2-parameter generalization of Poschl-Teller potential and the resulting 2d-Hamiltonians as generalizations of TTW and PW Hamiltonians on Euclidean plane \cite{l22} \cite{l22a} and 2d-spaces of constant curvature \cite{l25}, \cite{l28a}, \cite{l30}.\\ 
The paper is organized as follows. For a convenience of the reader as well as to fix the notation we start with brief recapitulation of the main ideas and results of Ref.\cite{l29}. Then, we explain how the implicit and transcendental, in general, equation on angular potentials can be solved in terms of elementary functions. We present and discuss the explicit solutions for both types of allowed radial potentials. Next we check in the independent way the superintegrability of considered systems. Finally we address the question of additional integrals of motion and we explicitly compute some of them. We conclude the paper with some remarks and open problems.

\section{Models}

\par

Models we shall be interested in are defined on a two-dimensional configuration space of constant curvature $k$. A positive-defined metric of the space when written in terms of geodesic polar coordinates $(r, \varphi ),$ reads

\begin{equation}
\label{r1}
ds^2 = dr^2 + s_k^2(r)d\varphi ^2,
\end{equation}
\;\\\
where the functions $s_k(r)$ are defined as follows:

\begin{equation}
\label{r2}
s_k(r) = 
\begin{cases}
\frac{1}{\sqrt{k}} sin(\sqrt{k}r) & k > 0\\
 r & k = 0\\
 \frac{1}{\sqrt{-k}} sh(\sqrt{-k}r) & k < 0\\
\end{cases}
\end{equation}
\;\\\
If $k=0\quad ds^2$ is the standard Euclidean metric of ordinary plane in the spherical coordinates. If $k$ is nonzero $ds^2$ represents the metric of sphere or hyperbolic plane for $k$ positive or negative, respectively.\\
The corresponding Hamiltonian separating in the $(r,\varphi )$ coordinates reads 
\;\\\
\begin{equation}
\label{r3}
H_{k}(r,\varphi , p_{r}, p_{\varphi }) = \frac{p_{r}^{2}}{2} + \frac{1}{s_k^2(r)}(\frac{p_{\varphi}^{2}}{2} + c(\varphi))+ a_k (r)
\end{equation}
\;\\\
where $(p_r,p_\varphi )$ stand for canonical momenta conjugated to $(r,\varphi )$ variables and
and $a_k (r)$, $c(\varphi)$  denote  radial and angular potentials, respectively.\\
Under rather general and mild assumptions on these potentials, separability of our Hamiltonian implies its integrability.\\
Involutive, global and functionally independent Liouville integrals are provided by a generalized "momentum" $l( \varphi ,p_{\varphi })$

\begin{equation}
\label {r4}
l( \varphi ,p_{\varphi }) = \frac{p_{\varphi}^{2}}{2} + c(\varphi) 
\end{equation}
\;\\
 and Hamiltonian itself\\
 \;\\\
\begin{equation}
\label{r5}
H(r,\varphi , p_{r}, p_{\varphi }) = \frac{p_{r}^{2}}{2} + \frac{1}{s_k^2(r)}l( \varphi ,p_{\varphi })+ a_k (r)
\end{equation}
\;\\\

By Arnold-Liouville theorem equations:\\
\begin{equation}
\label{r6}
\begin{split}
 H_k (r,\varphi , p_{r}, p_{\varphi }) = E\\ 
 l ( \varphi ,p_{\varphi }) = L 
 \end{split}
\end{equation}
\;\\\
define, for some intervals of the values of separation constants E and L a compact and sufficiently regular surface (which is isomorphic, by Arnold-Liouville theorem, to 2-d tori).\\
\par A superintegrability of the model (i.e. the existence of third globally defined and functionally independent constant of motion), unlike integrability itself, imposes strong conditions on the radial and angular potentials. These conditions can be  conveniently derived and studied
in terms of action-angle variables.\\
The relevant in our case angular $J_\varphi $ and radial $J^{k}_r$ actions are defined to be

\begin{equation}
\label{r7}
\begin{split}
J_{\varphi }(L) = \frac{1}{\pi }\int_{\varphi _{min}}^{\varphi _{max}}\sqrt{2(L - c(\varphi ))}d\varphi\\
\;\\\ 
J^{k}_{r}(E, L) = \frac{1}{\pi }\int_{r_ {min}}^{r _{max}}\sqrt{2(E- a^{k}(r) - \frac{L}{s_k^{2}(r)})}dr 
\end{split}
\end{equation}
\;\\\\
where $\varphi _{min}$, $\varphi _{max}$, $r_ {min}$, $r _{max}$ are the roots of the relevant integrands. \\
The corresponding angle variables $\psi _\varphi $, $\psi ^{k}_r$ are defined as 
\begin{equation}
\label{r8}
\begin{split}
\Psi^{k}_{r} = \frac{\partial{S^k}}{\partial{J^k_{r}}} \\
 \Psi_{\varphi } = \frac{\partial{S^k}}{\partial{J_{\varphi }}} 
\end{split} 
\end{equation}

where a generating function $S^k$ reads

\begin{equation}
\label{r9}
\begin{split}
& S^k = S^{k}(r,\varphi ,J^{k}_r, J_\varphi ) = S^k(r,\varphi ,E(\vec J), L(\vec J)) \\
\;\\\
& S^{k}(r,\varphi ,E, L) =  
\int_{\varphi _{0}}^{\varphi}\sqrt{2(L - U_\sigma (\varphi ))}d\varphi  + \int_{r_ 0}^{r }\sqrt{2(E- a^{k}(r) - \frac{L}{s_k^{2}(r)})}dr
\end{split} 
\end{equation}
\;\\
Now, it is known that the Hamiltonian of a maximally superintegrable system (2n-1 integrals of motion), when expressed in terms of the actions,
depends on their linear combination with integer coefficients. In our case we have

\begin{equation}
\label{r10}
H^k = H^k(\vec J) = H^k(mJ^{k}_r + nJ_{\varphi} )  \qquad m,n \in Z 
\end{equation}
\;\\\
It means that the linear combination of the actions is a function $f^k(E)$ of energy
\;\\\
\begin{equation}
\label{r11}
f^k(E) = mJ^{k}_{r}(E,L) + nJ_{\varphi}(L) 
\end{equation}
\;\\\
This has the following important consequences. A partial derivative of (\ref{r11}) with respect to L gives 
\;\\\ 
\begin{equation}
\label{r12}
m\frac{\partial J^k_{r}(E,L)}{\partial L} = - n\frac{\partial J_{\varphi}(L)}{\partial L} 
\end{equation}
\;\\\
which, in turn implies \\
\;\\\ 
\begin{equation}
\label{r13}
\frac{\partial}{\partial E}(\frac{\partial J^{k}_{r}(E,L)}{\partial L}) = 0
\end{equation}
\;\\\
The second of eqs.(\ref{r7}) allows one to write the partial derivative of the radial action with respect to L in the form
\;\\\
\;\\\
\begin{equation}
\label{r14}
\frac{\partial J^k_{r}}{\partial L} = \frac{-1}{\pi }\int_{r_{min}}^{r_{max}}
\frac{dr}{s_k^{2}(r)\sqrt{2(E- a^{k}(r) - \frac{L}{s_k^{2}(r)})} } 
\end{equation}
\;\\\
Hence, eq.(\ref{r13}) says that the integral on the R.H.S. of eq.(\ref{r14}) does not depend on the energy E. In fact, this is a condition on the radial potential $a^k(r)$. One can  show see Refs. \cite{l6}, \cite{l29} that it is satisfied if $a^k(r)$ is of oscillator type (it is usual harmonic oscillator potential on the plane for $k = 0$ and its spherical or hyperbolic counterpart for $k\neq 0$)
\;\\\ 
\;\\\
\begin{equation}
\label{r15}
 a^k(r) = 
\begin{cases}
 \gamma \mid k\mid \coth^2(\sqrt{-k}r) + \frac{\displaystyle \omega }{\displaystyle \mid k\mid }\frac{\displaystyle 1}{\displaystyle \coth^2(\sqrt{-k}r)} & \text{if}\;\;k\;<\;0\\
 \;\\
 \frac{\displaystyle \gamma}{\displaystyle r^2} + \omega  r^2 & \text{if}\;\;k\;=\;0\\
 \;\\
  \gamma \mid k\mid \cot^2(\sqrt{k}r) + \frac{\displaystyle \omega }{\displaystyle \mid k\mid }\frac{\displaystyle 1}{\displaystyle \cot^2(\sqrt{k}r)} & \text{if}\;\;k\;>\;0\\
\end{cases}
\end{equation}
\;\\\
\;\\\
or if $a^k(r)$ is of (generalized) Kepler type
\;\\\
\;\\\
\begin{equation}
\label{r16}
 a^k(r) = 
\begin{cases}
 B \mid k\mid \coth^2{(\sqrt{-k}r)} -  \sqrt{-k}\coth{(\sqrt{-k}r)}\sqrt{D + \mid k\mid F\coth^2{(\sqrt{-k}r)}} \\
 \;\\
 \frac{\displaystyle B}{\displaystyle r^2} - \frac{\displaystyle \sqrt{Dr^2 + F}}{\displaystyle r^2} \\
 \;\\
 B\mid k\mid \cot^2{(\sqrt{k}r)} -  \sqrt{k}\cot{(\sqrt{k}r)}\sqrt{D + \mid k\mid F\cot^2{(\sqrt{k}r)}} \\
\end{cases}
\end{equation}
\;\\
\;\\\
In the above formula the first line corresponds to $k<0$, the second one to $k=0$ while the last one to $k>0$.\\
In order to deal with physically interesting potentials( given by real, single valued, convex functions with a single minimum) we choose $\gamma ,\omega  , B, D, F \geq 0$ and $ B+ L +\sqrt{F} \geq 0 $ as well as $\gamma  + L\geq 0$. \\
For $F=0$ we have the ordinary Kepler potential on the plane ($k=0$) or its hyperbolic ($k<0$)/spherical ($k>0$) counterparts.\\
Now, passing to the angular potentials $c(\varphi )$ let us note that the first of
eqs.(\ref{r7}) allows to rewrite eq.(\ref{r12}) in the form of the Abel integral equation for the  potential $c(\varphi )$
\;\\ 
\begin{equation}
\label{r18}
-2\pi \frac{m}{n} \frac{\partial J^k_{r}(E, L)}{\partial L}= \sqrt{2}\int_{\varphi _{min}}^{\varphi _{max}}\frac{d\varphi }{\sqrt{(L - c(\varphi ))}}  
\end{equation}
\;\\\
The R.H.S. of the eq.(\ref{r18}) is nothing but a formula for a period $T(L)$ of recurrent trajectories with an "energy' $L$ in a bounded potential  $c(\varphi )$. Hence the angular potentials generating superintegrable dynamics can be described as isoperiodic ones i.e. the potentials with the same dependence $T(L)$ of the period $T$ on "energy"  $L$ 
\[
T(L) \equiv  -2\pi \frac{m}{n} \frac{\partial J^k_r(E,L)}{\partial L}
\]

Assuming $c(\varphi )$ to be a convex function on an interval $(\varphi _1, \varphi _2) \subset (0,2\pi )$ with a single minimum $c_0$ at $\varphi _0 \in (\varphi _1, \varphi _2)$ and two branches $c_{+} (\varphi )$ for $\varphi  > \varphi _0$ and $c_{-} (\varphi )$ for $\varphi  < \varphi _0$ both going to infinity as $\varphi$ approaches $\varphi _1$ or $\varphi _2$, respectively, the implicit solution to the Abel equation (\ref{r18}) reads \, \cite{l38}, \cite{l6}  
\;\\\
\begin{equation}
\label{r19}
\varphi _{\pm }(c) = \mp \frac{m}{n \sqrt{2}}\int_{c_{0}}^{c}\frac{\partial _{L}J^k_{r}(E, L)dL}{\sqrt{c - L}}  + G(c) 
\end{equation}
\;\\\
here,
$\varphi _{\pm }(c)$ denote the inverse maps of $c _{\pm }(\varphi)$, respectively and $G : R\rightarrow R$ is an arbitrary single valued function which does not spoil the assumed properties (a single valuedness, a single minimum at $\varphi _0$, convex character) of the potential $c(\varphi )$.\\
Knowing the radial potentials $a^k(r)$ given by eqs. (\ref{r15}) or (\ref{r16}) one can find an explicit formula on partial derivatives of the radial actions with respect to $L$ (this can be done directly by performing integral (\ref{r14}) or by computing the relevant actions first and then taking  the derivatives)\\
\;\\ 
\begin{equation}
\label{r17}
 \frac{\partial J^k_{r}(E, L)}{\partial L} = 
\begin{cases}
 -\frac{\displaystyle 1}{\displaystyle 2\sqrt{2}}\displaystyle{\frac{1}{\sqrt{L + \gamma }}}  & \text{oscillator type }\\
 \;\ \\
 -\frac{\displaystyle 1}{\displaystyle 2\sqrt{2}}\left ( \displaystyle{\frac{1}{\sqrt{L+B + \sqrt{F}}} + \frac{1}{\sqrt{L+B - \sqrt{F}}}}\right ) &  \text{(gen.)Kepler type} \\
\end{cases}
\end{equation}
\;\\
\;\\
Note that in the case of central potentials i.e. for $c(\varphi) = 0$ (\ref{r17}) implies the eq. (\ref{r18}) holds identically for any $L$ provided  $\gamma = 0$, $m/n = 2$ for oscillator radial potentials and $B = 0 =F$, $m/n = 1$ for generalized Kepler ones. This is nothing but Bertrand theorem.\\
 Now, computing the integral on the R.H.S. of eq.(\ref{r19}) with $\partial _{L}J^k_{r}(E, L)$ given by eq.(\ref{r17}) we obtain a general implicit formula
$\varphi  = \varphi (c)$ for angular potentials leading to superintegrable dynamics
\\
\begin{equation}
\label{r20}
 \varphi _{\pm }(c) = \pm \frac{1}{2\nu }(\frac{\pi }{2} - \arcsin{f(c)}) + G(c) 
\end{equation}
\;\\
 where the function $f(c)$ is given by
;\\ 
\;\\
\begin{equation}
\label{r21}
 f(c) = 
\begin{cases}
 \frac{\displaystyle 2c_0 +\gamma - c}{\displaystyle c + \gamma }  & \text{for $a^k(r)$ of oscillator type }\\
 \;\ \\
 \frac{\displaystyle c_0 + \sqrt{(c_{0} + B)^2 - F} - c}{\displaystyle \sqrt{(c  + B)^2 - F}} &  \text{for $a^k(r)$ of (generalized)Kepler type} \\
\end{cases}
\end{equation}
\;\\
\;\\
For $F=0$ i.e. for Kepler potential $a^k(r)$ the function $f(c)$ takes the form
\;\\
\;\\
\begin{equation}
\label{r22}
f(c) = \frac{\displaystyle 2c_0 + B - c}{\displaystyle c + B }
\end{equation}
\;\\
\;\\
The parameter $\nu $ in eq.(\ref{r20}) is given by
\\ 
\;\\
\begin{equation}
\label{r23}
 \nu  = 
\begin{cases}
 \frac{\displaystyle 2n}{\displaystyle m}  & \text{for $a^k(r)$ of oscillator type }\\
 \;\ \\
 \frac{\displaystyle n}{\displaystyle m} &  \text{for $a^k(r)$ of (generalized)Kepler type} \\
\end{cases}
\end{equation}
\;\\
\;\\
\section{The explicit form of angular potentials}

\begin{figure}[th!]
\begin{center}
\begin{tikzpicture}[scale=0.7]
\draw[thick, fill=yellow!50] (-2.5cm, 0) -- (0, 2.5cm) -- (2.5cm, 0) -- (0,-2.5cm) -- cycle;
\draw[->] (0,-4cm)  -- (0, 4cm) node [anchor=north west]{$\alpha$};
\draw[->] (-4cm, 0)  -- (4cm, 0) node [anchor=north east]{$\beta$};
\draw[very thin] (-2.5cm, -0.15cm) node [anchor=north]{-1} -- (-2.5cm, 0.15cm);
\draw[very thin] (2.5cm, -0.15cm) node [anchor=north]{1} -- (2.5cm, 0.15cm);
\draw[very thin] (-0.15cm, -2.5cm) node [anchor=east]{-1} -- (0.15cm, -2.5cm);
\draw[very thin] (-0.15cm, 2.5cm) node [anchor=east]{1} -- (0.15cm, 2.5cm);
\end{tikzpicture}
\end{center}
\caption{The set of $\beta \alpha $ parameters defining angular potentials $c(\varphi )$
 leading to a superintegrable dynamics}
\label{kwadrat}
\end{figure}
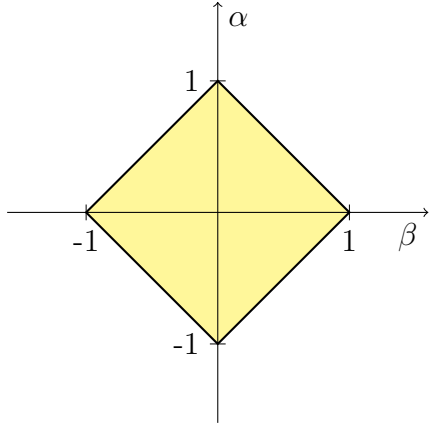

For an arbitrary allowed function $G(c)$ equation (\ref{r20}) is, in general, the transcendental one admitting no solutions for $c = c(\varphi )$ in terms of elementary functions, although such solutions exists for particular choices of G. For example, taking $G(c) = const = \varphi _0$ results in   (considered in  Ref.( \cite{l6}))   potentials $c(\varphi ) = f^{-1}(cos2\nu (\varphi  - \varphi _0))$ with a function $f^{-1}$ given by eq.(\ref{r26}) or (\ref{r27}) below.\\
We aim to find another functions $G$ generating the angular potentials $C(\varphi )$ given in terms of elementary functions.\\
A possible strategy is to write equation (\ref{r20}) with $f(c)$ given by eq. (\ref{r21}) in the form
\;\\
\;\\
\begin{equation}
\label{r24}
\cos{2\nu (\varphi _{\pm}  - G(c))} = f(c)
\end{equation}
\;\\
\;\\
and then to consider it as a system of two equations
\;\\
\begin{equation}
\label{r25}
\begin{cases}
 f(c) = \tilde f\\
  \cos{2\nu (\varphi _{\pm}  - G(c))} = \tilde f\\
\end{cases}
\end{equation}
\;\\
\;\\
As $f(c) \in (-1, 1>$,  $f(c_0) = 1$, and $f(c) \longrightarrow  -1$ for $c \longrightarrow  \infty $ we have\\
$\tilde f \in (-1, 1>$,  $\tilde f_0\equiv  f(c_0) = 1$, and $\tilde f = f(c) \longrightarrow  -1$ for $c \longrightarrow  \infty $.
\;\\
\;\\
Solving first equation of the system  (\ref{r25}) with respect to $c$ gives a function $c(\tilde f ) = f^{-1}(\tilde f)$
\;\\
\begin{equation}
\label{r26}
c(\tilde f) = \frac{\displaystyle 2(\gamma  + c_0)}{\displaystyle \tilde f + 1} - \gamma 
\end{equation}
\;\\
for  the radial potentials of the oscillator type and\\
\;\\ 
\begin{equation}
\label{r27}
c_\pm (\tilde f)=
\begin{cases}
\frac{\displaystyle J + B\tilde f^2 \pm  \tilde f \sqrt{(J + B)^2 + F(\tilde f^2 - 1)}}{\displaystyle 1 - \tilde f^2} & \tilde f \neq -1,1\\
\;\\
c_0 & \tilde f = 1
\end{cases}
\end{equation}
\;\\
where $J\equiv c_0 + \sqrt{(c_{0} + B)^2 - F}$ for the radial potentials of (generalized)Kepler type .\\
Rewriting eq.(\ref{r27}) in an equivalent way as
\;\\
\begin{equation}
\label{r28}
c\pm (\tilde f) = \frac{\displaystyle J^2 + (F - B^2)\tilde f^2}{\displaystyle J + B\tilde f^2 \mp   \tilde f \sqrt{(J + B)^2 + F(\tilde f^2 - 1)}} 
\end{equation}
\;\\
shows that $c_\pm (\tilde f)$  have no singularity at $\tilde f = 1$.\\
For the Kepler radial potential ($F=0$)  $c(\tilde f)$ is given by eq.(\ref{r26}) with $\gamma $ replaced by $B$.\\
Note that neither $c(\tilde f)$ nor $c_{\pm} (\tilde f)$ have local minimum in the interval $ (-1,1>$ and consistently with the relations below the equation (\ref{r25}) $c(\tilde f_0 = 1) = c_0$ and $c(\tilde f\rightarrow -1)\longrightarrow \infty $. Similarly, $c_{\pm} (\tilde f_0 = 1) = c_0$ and
$c_{-}(\tilde f\rightarrow -1)\longrightarrow \infty $ for $J + B > 0$ while $c_{+}(\tilde f\rightarrow -1)\longrightarrow \infty $ for $J + B < 0$.
This shows that $c_{-}(\tilde f)$ is relevant in $J + B > 0$ case while $c_{+}(\tilde f)$ is relevant in $J + B < 0$ one and that $\tilde f = -1$, $\tilde f = 1$ correspond to the boundaries of domains of the potentials $c(\tilde f (\varphi ))$, $c_{\pm} (\tilde f(\varphi ))$ and their minima, respectively. \\
Having determined the functions $c(\tilde f ) = f^{-1}(\tilde f)$ we rewrite the second equation of the system  (\ref{r25}) in the form 
\;\\
\begin{equation}
\label{r29}
  \cos{2\nu (\varphi - \tilde G(\tilde f))} = \tilde f\\
\end{equation}
\;\\
where $\tilde G(\tilde f)) = G(f^{-1}(\tilde f))$. Then we look for solutions $\tilde f = \tilde f(\varphi )$ to eq.(\ref{r29}) which can be written in terms of elementary functions and once composed with functions $ c = c(\tilde f )$ given by eqs. (\ref{r26}) or (\ref{r27}) provide potentials $c(\varphi  ) = f^{-1}(\tilde f(\varphi ))$  enjoying the required properties.\\
To this end we use basic trigonometric identities and algebraic operations to write eq.(\ref{r29}) in the form 
\;\\
\begin{equation}
\label{r30}
\tilde f^2 - 2\tilde f\cos{(2\nu \tilde G)}\cos{2\nu \varphi }+ \cos^2{(2\nu \tilde G)} - \sin^2{2\nu \varphi}  = 0
\end{equation}
\;\\
Putting now \\
\;\\
\begin{equation}
\label{r31}
\cos{2\nu \tilde G (\tilde f)} = \alpha \tilde f + \beta   \qquad       \alpha ,\beta  \in  R
\end{equation}
\;\\
we get\\
\;\\
\begin{equation}
\label{r32}
a(\varphi )\tilde f^2 + b(\varphi )\tilde f + d(\varphi ) = 0
\end{equation}
\;\\
where
\;\\
\begin{equation}
\label{r33}
\begin{cases}
a(\varphi ) = (\cos{2\nu \varphi}  - \alpha )^2 + \sin^{2}{2\nu \varphi}  = 1 + \alpha ^2 - 2\alpha \cos{2\nu \varphi }\\
\;\\
b(\varphi ) = -2\beta (\cos{2\nu \varphi}  - \alpha )\\
\;\\
d(\varphi ) = \beta ^2 - \sin^{2}{2\nu \varphi}
\end{cases}
\end{equation}
\;\\
\;\\
Obviously, the ansatz (\ref{r31}) is consistent provided that for every $\tilde f \in  <-1, 1>$ $\mid \alpha \tilde f + \beta  \mid \leq 1$ which in turn implies\\
\;\\
\begin{equation}
\label{r34}
\mid\alpha  \mid + \mid \beta \mid   \leq  1
\end{equation}
\;\\
In other words $\alpha ,\beta $ parameters live inside and on the square $\mid\alpha  \mid + \mid \beta \mid   =  1$ displayed in the Figure~\ref{kwadrat}.

\begin{figure}
\begin{center}
\begin{tikzpicture}[xscale=2, yscale=0.5]
\begin{scope}[yscale=3]
\draw[->] (-0.25, 0) -- (6, 0) node[anchor=north east]{$\phi$};
\draw[->] (0, -1.2) -- (0,1.6) node[anchor=north east]{$\tilde f_{-}(\phi)$};
\draw[dashed, blue] (-0.1,1) node[anchor=east]{1} -- (6, 1);
\draw[dashed, blue] (-0.1,-1) node[anchor=east]{-1} -- (6, -1);
\draw[dashed, blue] (5.1051, -1.2) -- (5.1051, 0) node[anchor=north west]{$\frac{13}{16}\pi$}
    --  (5.1051, 6);
\draw[dashed, blue] (0.3927, -1.2) -- (0.3927, 0) node[anchor=north west]{$\frac{1}{16}\pi$}
    --  (0.3927, 6);
\draw[gray, dashed, very thin] (1.36186, -1.2) node[anchor= east]{$\phi_{\text{min}}$}
-- (1.36186, 0) 
-- (1.36186, 6) ;
\draw[gray, dashed, very thin] (4.72835, -1.2) 
node[anchor=east]{$\phi_{\text{max}}$} -- (4.72835, 0)
   -- (4.72835, 6);
\draw[gray, dashed, very thin] (-0.1, -0.62717) node[anchor=east]{$f_L$} -- (6, -0.62717);
\draw[gray, dashed, very thin] (3.92904, -1.2) 
node[anchor=east]{$\phi_{0}$} 
-- (3.92904, 0)
  -- (3.92904, 6);
\draw[red, thick] plot[smooth] file{f2.3.table}; 
\end{scope}
\begin{scope}[yshift=8cm]
\draw[->] (-0.5, 0) -- (6, 0) node[anchor=north east]{$\phi$};
\draw[->] (0, -1.5) -- (0,10) node[anchor=north east]{$c(\phi)$};
\draw[dashed, blue] (-0.1,-1) node[anchor=east]{-1} -- (6, -1);
\draw[dashed, blue] (-0.1,5) node[anchor=east]{5} -- (6, 5);
\draw[red, thick] plot[smooth] file{c2.3.table};
\draw[gray, dashed, very thin] (-0.1, 7.59687) node[anchor=east]{$L$} -- (6, 7.59687);
\end{scope}
\end{tikzpicture}
\end{center}
\caption{The relation between the potential $c(\varphi )\equiv f^{-1}(\tilde f(\varphi ))$ and the function $\tilde f=\tilde f(\varphi ))$; $\tilde f(\varphi )$, $c(\varphi)$ are given by eqs (\ref{r35}), (\ref{r37}), respectively for $\alpha = \frac{1+\sqrt{3}}{2}$, $\beta  = \frac{1-\sqrt{3}}{2}$ , $\gamma =3$, $c_0 =-1$, $\nu =\frac{4}{3}$. }
\label{duza}
\end{figure}
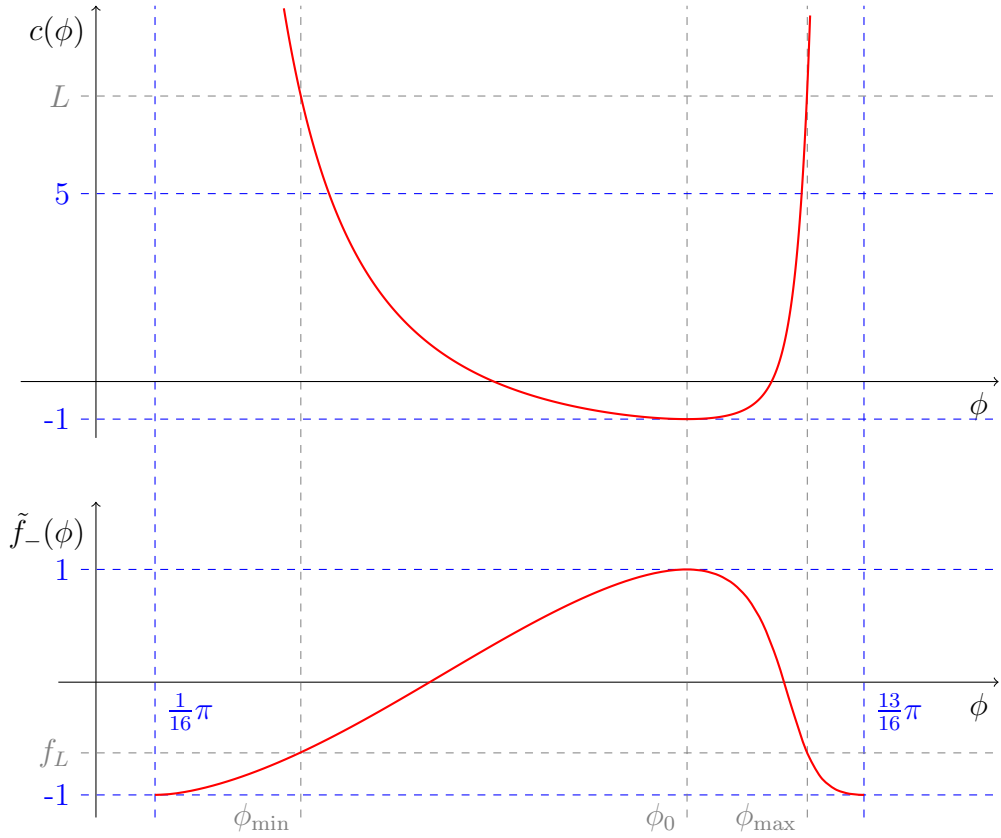

It is remarkable that the condition (\ref{r34}) appears to be sufficient for solutions of eq.(\ref{r32}) to generate angular potentials $c(\varphi )$
enjoying required properties and leading to a superintegrable dynamics.\\
In fact, the condition (\ref{r34}) implies that discriminant $\Delta  = 4\sin^2{2\nu \varphi} (a(\varphi ) - \beta^2 )$ of quadratic equation (\ref{r32}) is not negative. Consequently, for any $\varphi $ we have  two real solutions $\tilde f_\pm(\varphi ) $ to the eq.(\ref{r32}) (which reduce to the one at points $2\nu \varphi  = k\pi $ where $\Delta  = 0$)
\;\\
\begin{equation}
\label{r35}
\tilde f_{\pm }(\varphi ) = \frac{\beta (\cos{2\nu \varphi}  - \alpha ) \pm \sin{2\nu \varphi }\sqrt{a(\varphi ) - \beta ^2}}{a(\varphi )}
\end{equation}
\;\\
where $a(\varphi )$ is defined by the first of the eqs.(\ref{r33}).
\;\\
\;\\
Both solutions $\tilde f_{\pm }(\varphi )$ (corresponding to $ \tilde G (\tilde f) = \pm \frac{\displaystyle 1}{\displaystyle 2\nu }\arccos{(\alpha \tilde f + \beta)} $, respectively) are related by the transformation $2\nu \varphi  \longrightarrow  2\pi  - 2\nu \varphi$. Therefore, without loosing generality we shall use $\tilde f_{-}(\varphi )$, denoted as $\tilde f(\varphi )$, henceforth.\\
The function $\tilde f(\varphi )$ is periodic with the period $T = \frac{\pi }{\nu }$ and has the following properties
\;\\
\begin{equation}
\label{r36}
\begin{split}
\tilde f(\tilde \varphi ) = -1 \quad \text{iff} \quad \cos{2\nu \tilde \varphi}  = \alpha -  \beta \quad \sin{2\nu \tilde \varphi} = \sqrt{1 - (\alpha -  \beta)^2}\\
\;\\
\tilde f( \varphi_0 ) = 1 \quad \text{iff} \quad \cos{2\nu \varphi_0} =  \alpha + \beta \quad \sin{2\nu \varphi_0} = -\sqrt{1 - (\alpha + \beta)^2}\\
\end{split} 
\end{equation}
\;\\
\;\\
As explained below the formula (\ref{r28}) the first equation  (\ref{r36}) determines the interval\\
\begin{equation}
\label{r36a}
\varphi \in (\tilde \varphi ,\tilde \varphi +\frac{\pi }{\nu })\equiv (\frac{\displaystyle 1}{\displaystyle 2\nu }\arccos{(\alpha - \beta)}, \frac{\displaystyle 1}{\displaystyle 2\nu }\arccos{(\alpha - \beta)} +   \frac{\displaystyle \pi }{\displaystyle \nu })
\end{equation}
 where the potential  $c(\varphi  ) = f^{-1}(\tilde f(\varphi ))$ is defined, while the second one gives a point $\varphi _0$ where the unique maximum of the $\tilde f(\varphi )$ in this interval i.e. the unique minimum of the potential $c(\varphi )$ is attained.
Indeed, one can check that other points where the first derivative of $\tilde f(\varphi )$ vanishes do not belong to the interval (\ref{r36a})
if condition (\ref{r34}) is satisfied.\\
Due to these properties composing the functions $\tilde f = \tilde f(\varphi )$ and 
$c(\tilde f ) = f^{-1}(\tilde f)$ ones given by eqs. (\ref{r26}), (\ref{r27}), results in the angular potentials 
$c(\varphi ) = c(\tilde f(\varphi ))$  generating the superintegrable dynamics.\\
The relation between the potential $c(\varphi )\equiv f^{-1}(\tilde f(\varphi ))$ and the function $\tilde f=\tilde f(\varphi ))$ illustrates a Figure~\ref{duza}.

It is worth  emphasizing, that in spite of being described, in general, by complicated formulas the potentials behave in a very simple way. They have one single minimum and go to infinity at the ends of intervals they are defined on.\\
Below we present some explicit formulas for $c(\varphi )$. \\
We start with angular potentials corresponding to the radial oscillator/Kepler type ones. For the parameters $\alpha ,\beta $ inside of the square $|\alpha | + |\beta | = 1$ (i.e. for $\alpha ,\beta $ satisfying $|\alpha | + |\beta | < 1$) inserting eq. (\ref{r35}) into eq.(\ref{r26}) generates a potential
\;\\
\;\\
\begin{equation}
\label{r37}
c(\varphi ) = \frac{2(\gamma +c_0)a(\varphi )}{\beta (\cos{2\nu \varphi } -\alpha ) - \sin{2\nu \varphi }\sqrt{a(\varphi ) -\beta ^2} + a(\varphi )} -\gamma 
\end{equation}
\;\\
\;\\
here $\varphi $ belongs to the interval (\ref{r36a}), while $c_0 = c(\varphi _0)$
where $\varphi _0 = \frac{\displaystyle \arccos{(\alpha +\beta )}}{\displaystyle 2\nu }$, $\quad \sin{2\nu \varphi_0} =-\sqrt{1-(\alpha +\beta )^2}$  (see second equation(\ref{r36}))  is the single minimum of the $c(\varphi )$ in this interval. \\
The above general formula (\ref{r37}) simplifies for parameters $\alpha$ , $\beta $ living on the boundary $|\alpha | + |\beta | = 1$.\\ 
In particular, in a second quadrant  of $\beta \alpha $ plane i.e. for $\alpha \in (0,1)$ and $\beta =\alpha -1$, the function (\ref{r35}) reads
\;\\
\;\\
\begin{equation}
\label{r38}
\tilde f(\varphi ) \equiv \tilde f^{II}(\varphi ) =  = \frac{(\alpha -1)(\cos{2\nu \varphi}  - \alpha ) -2\sqrt{\alpha } \sin{2\nu \varphi } \mid \sin{\nu \varphi }\mid }{1 + \alpha ^2 -2\alpha \cos{2\nu \varphi }}
\end{equation}
\;\\
\;\\
and generates via eq. (\ref{r26}) a potential
\;\\
\;\\
\begin{equation}
\label{r39}
c^{II}(\varphi ) = (\gamma +c_0)\frac{(\cos{\nu \varphi} + \sqrt{\alpha })^2}{\sin ^2 {\nu \varphi }} + c_0
\end{equation}
\;\\
\;\\
here $ \nu \varphi \in (0,\pi )$ and $c_0=c(\varphi _0)$ is the unique minimum of $c$ corresponding to the angle $\nu \varphi _0=\arccos {\sqrt{\alpha }}\quad$ ($\cos{2\nu \varphi _0} = 2\alpha -1$ and $\sin{2\nu \varphi _0} = -2\sqrt{\alpha }\sqrt{(1-\alpha )}$).\\
The potential $c^{II}$ is nothing but the celebrated Poschl-Teller potential. Indeed, in terms of $\frac{\displaystyle \nu \varphi }{\displaystyle 2}$ angels $c^{II}(\varphi )$ can be written in more familiar form as
\;\\
\;\\
\begin{equation}
\label{r40}
c^{II}(\varphi )= \frac{A_-}{\cos^2{\frac{\nu \varphi }{2}}} + \frac{A_+}{\sin^2{\frac{\nu \varphi }{2}}} - \gamma 
\end{equation}
\;\\
where
\;\\
\begin{equation}
\label{r41}
A_{\pm} = \frac{\gamma + c_0}{4} (1 \pm \sqrt{\alpha })^2
\end{equation}
\;\\
\;\\
If angular potential is given by $c^{II}(\varphi )$ and the radial potential $a^k(r)$ is that of oscillator type ($\nu =2\frac{\displaystyle n}{\displaystyle m}$)
we are dealing with TTW model or its generalization to the curved space. For the $a^k(r)$ of Kepler type ($\nu =\frac{\displaystyle n}{\displaystyle m}$, $\gamma \rightarrow B$)
we get P-W model in the plane or curved space. \\
Similarly, in the fourth quadrant of $\beta \alpha $ plane i.e. for $\alpha \in (-1,0)$ and $\beta =\alpha +1$  we obtain the function
\;\\
\;\\
\begin{equation}
\label{r42}
\tilde f(\varphi ) \equiv \tilde f^{IV}(\varphi ) =   \frac{(\alpha + 1)(\cos{2\nu \varphi}  - \alpha ) -2\sqrt{\alpha } \sin{2\nu \varphi } \mid \cos{\nu \varphi }\mid}{1 + \alpha ^2 -2\alpha \cos{2\nu \varphi }}
\end{equation}
\;\\
\;\\
and the related potential
\;\\
\;\\
\begin{equation}
\label{r43}
\begin{split}
c^{IV}(\varphi ) = (\gamma +c_0)\frac{(\sin{\nu \varphi} - \sqrt{\alpha })^2}{\cos ^2 {\nu \varphi }} + c_0=\\
= \frac{A_-}{\cos^2{(\frac{\nu \varphi }{2} +\frac{\pi }{4})}} + \frac{A_+}{\sin^2{(\frac{\nu \varphi }{2} +\frac{\pi}{4 })}} - \gamma
\end{split}
\end{equation}
\;\\
\;\\
where $ \nu \varphi \in (\frac{\pi}{2},\frac{3}{2}\pi )$ and the single minimum $c_0=c^{IV}(\varphi _0)$  corresponds to the angle $\nu \varphi _0=\pi -\arccos {\sqrt{1 +\alpha }}\quad$ ($\cos{2\nu \varphi _0} = 2\alpha +1$ and $\sin{2\nu \varphi _0} = -2\sqrt{-\alpha }\sqrt{(1+\alpha )}$). The constants $A_\pm$ are given by eqs.(\ref{r41}).\\
On the other hand for parameters $\alpha ,\beta $ from the first or third quadrant of $\beta \alpha $ plane we get the following functions $\tilde f$ and potentials $c(\varphi )$.\\
In the first quadrant that is when $\alpha \in (0, 1)$ and $\beta = 1 - \alpha $ 
\;\\
\;\\
\begin{equation}
\label{r44}
\tilde f(\varphi ) \equiv \tilde f^{I}(\varphi ) =  \frac{(1 - \alpha )(\cos{2\nu \varphi}  - \alpha ) -2\sqrt{\alpha } \sin{2\nu \varphi } \mid \sin{\nu \varphi }\mid }{1 + \alpha ^2 -2\alpha \cos{2\nu \varphi }}
\end{equation}
\;\\
\;\\
the corresponding potential reads\\
\;\\
\;\\
\begin{equation}
\label{r45}
c^{I}(\varphi ) =
\begin{cases}
(\gamma +c_0)\frac{\displaystyle \sin^2{\nu \varphi}}{\displaystyle (\cos{\nu \varphi } - \sqrt{\alpha })^2} + c_0 & \nu  \varphi \in (\arccos {\sqrt{\alpha }},\pi )  \\
\;\\
(\gamma +c_0)\frac{\displaystyle \sin^2{\nu \varphi}}{\displaystyle (\cos{\nu \varphi } + \sqrt{\alpha })^2} + c_0 & \nu  \varphi \in (\pi ,\arccos {\sqrt{\alpha }} +\pi )  
\end{cases}
\end{equation}
\;\\
\;\\
In these formulas $ \nu  \varphi \in (\arccos {\sqrt{\alpha }}, \arccos {\sqrt{\alpha }} +\pi)$ and the unique minimum $c_0=c^{I}(\varphi _0 )$ is at $\nu \varphi _0=\pi$.\\
Finally, in the third quadrant of $\beta \alpha $  plane i.e. for  $\alpha \in (-1, 0)$ and $\beta = -( 1 + \alpha) $ the function $\tilde f$ is given by
\;\\
\;\\
\begin{equation}
\label{r46}
\tilde f(\varphi ) \equiv \tilde f^{III}(\varphi ) =  = \frac{-(\alpha +1)(\cos{2\nu \varphi}  - \alpha ) -2\sqrt{-\alpha } \sin{2\nu \varphi } \mid \cos{\nu \varphi }\mid }{1 + \alpha ^2 -2\alpha \cos{2\nu \varphi }}
\end{equation}
\;\\
\;\\
and it leads to the potential\\
\;\\
\;\\
\begin{equation}
\label{r47}
c^{III}(\varphi ) =
\begin{cases}
(\gamma +c_0)\frac{\displaystyle \cos^2{\nu \varphi}}{\displaystyle (\sin{\nu \varphi } - \sqrt{\alpha })^2} + c_0 & \nu  \varphi \in (\arccos {\sqrt{\alpha + 1}},\frac{\pi}{2} )  \\
\;\\
(\gamma +c_0)\frac{\displaystyle \sin^2{\nu \varphi}}{\displaystyle (\sin{\nu \varphi } + \sqrt{\alpha })^2} + c_0 & \nu  \varphi \in (\frac{\pi}{2} ,\arccos {\sqrt{\alpha +1 }} +\pi )  
\end{cases}
\end{equation}
\;\\
\;\\
In this case $\nu  \varphi \in (\arccos {\sqrt{\alpha + 1}},\arccos {\sqrt{\alpha +1 }} +\pi ) $ while the single minimum $c_0=c^{III}(\varphi _0 )$ is
attained at $\nu \varphi _0 = \frac{\pi}{2}$.\\
The two branches  of the potential $c^{I}(\varphi )$ ($c^{III}(\varphi )$) are described by slightly different formulas. This is due to the fact that in the equation for the function $\tilde f^{I}(\varphi )$ ($\tilde f^{III}(\varphi )$) there is the absolute value of $\sin (\cos )$ function which changes a sign on the domain of $\tilde f^{I}(\varphi )$ ($\tilde f^{III}(\varphi )$). Nevertheless, the $c^{I}(\varphi )$ ($c^{III}(\varphi )$) potential is given by analytical smooth function. Note that here, contrary to the Poschl-Teller potential the coordinate of the minimum is fixed and independent of the parameter $\alpha $ which controls here the ends of the domain of the potentials. \\
Regarding the models with radial potentials $a^k(r)$ of generalized Kepler type given by eq.(\ref{r16}) we note that direct inserting of the eq.(\ref{r35}) into the eq.(\ref{r27}) would result in a very complicated general formula for the corresponding angular potential $c(\varphi )$ not especially useful for studying  its structure and properties. However, there is a choice of parameters $\alpha ,\beta $ providing relatively simple expressions on angular potentials corresponding to both type  (oscillator and generalized Kepler) of radial potentials. Indeed, putting $\beta  = 0$ in the eq.(\ref{r35}) we obtain      
\;\\
\begin{equation}
\label{r48}
\tilde f(\varphi ) = \frac{- \sin{2\nu \varphi }}{\sqrt{\tilde a(\varphi )}} 
\end{equation}
\;\\
$\tilde a(\varphi )$ is defined in first of eq.(\ref{r33}). The function (\ref{r48}) is defined on the interval $(\nu \tilde \varphi ,\nu \tilde \varphi + \pi )$, where $\cos{2\nu \tilde \varphi } = \alpha $ and $\sin{2\nu \tilde \varphi } = \sqrt{1 - \alpha^2}$. It takes a value one at a point $\nu \varphi _0$ where $\cos{2\nu \varphi_0 } = \alpha $ and $\sin{2\nu \varphi_0 } = - \sqrt{1 - \alpha^2}$.\\
The angular potential yielded by the function (\ref{r48}) and corresponding to the radial potential of oscillator/Kepler type reads
\;\\
\;\\
\begin{equation}
\label{r49}
\begin{split}
& c(\varphi ) =(\gamma + c_0)\frac{\sqrt{a(\varphi )} + \sin{2\nu \varphi }}{\sqrt{a(\varphi )} - \sin{2\nu \varphi }} +c_0=\\
&= 2(\gamma +c_0) \frac{\sin^2{2\nu \varphi } + \sin{2\nu \varphi }\sqrt{\sin^2{2\nu \varphi } + (\cos{2\nu \varphi } - \alpha )^2}}
{(\cos{2\nu \varphi} - \alpha )^2} +\gamma + 2c_0
\end{split}
\end{equation}
\;\\
\;\\
while the $c(\varphi )$ relevant for the radial potential of general Kepler type is given by
\;\\
\begin{equation}
\label{r50}
c_-(\varphi ) = \frac{(J+B)\sin^2(2\nu \varphi ) +\sin{2\nu \varphi }\sqrt{\big ( (J+B)^2 -F \big ) (\cos{2\nu \varphi } -\alpha )^2 + (J+B)^2\sin^2{2\nu \varphi }}}{(\cos{2\nu \varphi} - \alpha )^2} + J
\end{equation}
\;\\
\;\\
Both potentials (\ref{r49}) end (\ref{r50}) are defined on the interval $(\nu \tilde \phi ,\nu \tilde \varphi + \pi )$, and attain a single minimum $c_0 = c(\nu \varphi _0)$.\\
The plot illustrating the potential (\ref{r50}) is shown in the Figure 3.

\begin{figure}
\begin{center}
\begin{tikzpicture}[xscale=1.9, yscale=0.5]
\draw[->] (-0.5, 0) -- (6,0) node[anchor=north east]{$\phi$};
\draw[->] (0, -0.5) -- (0,11) node[anchor=north east]{$c_{-}(\phi)$};
\draw[dashed, blue] (-0.1, 8) node[anchor=east]{8} -- (6, 8);
\draw[dashed, blue] (0.7854, -0.1) node[anchor=north]{$\frac{\pi}{4}$} -- (0.7854, 11);
\draw[dashed, blue] (5.4978, -0.1) node[anchor=north]{$\frac{7}{4}\pi$} --  (5.4978, 11);
\draw[red, thick] plot[smooth] file{cc2.3.table}; 
\end{tikzpicture}
\end{center}
\caption{The plot of potential (\ref{r50}) for $\alpha = \frac{1}{2}$, $B = 1$ , $J =1$, $F =\frac{1}{2}$, $\nu =\frac{2}{3}$.}
\end{figure}
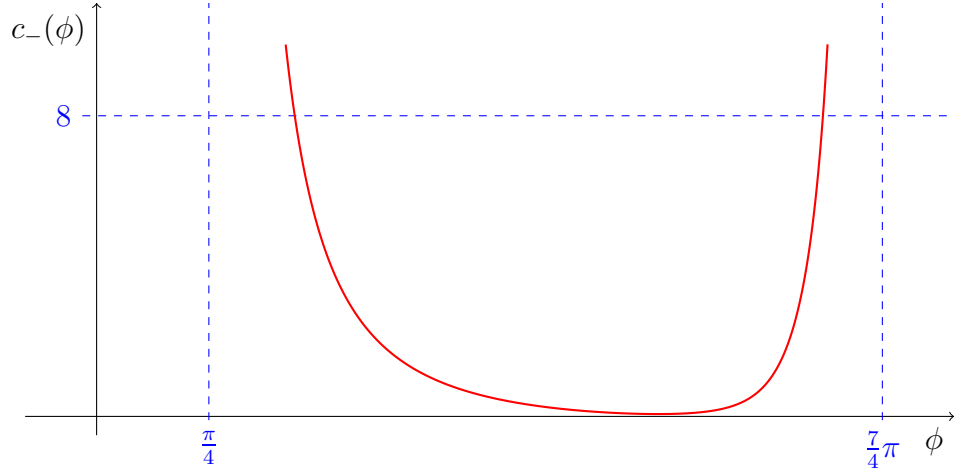

\section{Superintegrability}

Now, the superintegrability of the models defined by the Hamiltonian(\ref{r3}) (cf.eqs. (\ref{r15}) and (\ref{r16}) for radial potentials and (\ref{r35}), (\ref{r26}) and (\ref{r27})
for angular ones) 
can be independently confirmed by computing explicitly the radial and angular actions and checking that the relevant Hamiltonian depends on linear combinations of these actions with integer coefficients.\\
The integrals defining the radial actions (\ref{r7}) are elementary, performing them we get
\;\\
\begin{equation}
\label{r51}
J^k_r(E,L) = g^k(E) - \frac{\sqrt{2}}{2} \sqrt{L+\gamma }
\end{equation}
\;\\
for the potentials $a^k(r)$ of oscillator type given by eq.(\ref{r15}).\\
The $g^k(E)$ denote corresponding energy functions. There is no need to write them explicitly here.

Similarly, for generalized radial potential of Kepler type given by equation (\ref{r16}) we have
\;\\
\begin{equation}
\label{r52}
J^k_r(E,L) = h^k(E) - \frac{\sqrt {2}}{2}\Big ( \sqrt {L + B + \sqrt{F}} + \sqrt {L+B-\sqrt {F}}\Big )
\end{equation}
\;\\
where as above $h^k(E)$ are energy functions related to the relevant $a^k(r)$ radial potentials (\ref{r16}).\\
To compute the angular actions (given by first of equations (\ref{r7})) it is convenient to calculate first their partial derivatives $\frac{\partial J_\varphi }{\partial L}$ with respect to $L$. Then taking into account that $J_\varphi (L = c_0) = 0$  one finds the relevant actions.\\
To perform, in general, non elementary integrals\\
\;\\ 
\begin{equation}
\label{r53}
\int_{\varphi _{min}}^{\varphi _{max}}\frac{d\varphi }{\sqrt{L - c(\varphi )}} \equiv \int_{\varphi _{min}}^{\varphi _{max}}\frac{d\varphi }{\sqrt{L - c(\tilde f (\varphi ))}} 
\end{equation}
\;\\
determining the derivatives $\frac{\partial J_\varphi }{\partial L}$ we substitute\\
\;\\
\begin{equation}
\label{r54}
\varphi  = \varphi (\tilde f) =
\begin{cases}
\varphi_+ (\tilde f) \qquad for \qquad \varphi \geq \varphi _0
\;\\
\varphi_- (\tilde f) \qquad for  \qquad \varphi < \varphi _0
\end{cases}
\end{equation}
\;\\  
where\\
\;\\
$2\nu \varphi_\pm  (\tilde f) = \pm \arccos{\tilde f} - \arccos{(\alpha \tilde f + \beta )}$\\
\;\\
and $\nu  = \frac{2n}{m}$ for  $c(\varphi )$ corresponding to the radial $a^k(r)$ potentials of oscillator type, while $\nu =\frac{n}{m}$ for the case of $c(\varphi )$ related to $a^k(r)$ of (generalized) Kepler type.\\

Consequently, the differential reads\\
\;\\
\begin{equation}
\label{r55}
d\varphi  = 
\begin{cases}
d\varphi_+ \qquad for \qquad \varphi \geq \varphi _0
\;\\
d\varphi_- \qquad for  \qquad \varphi < \varphi _0
\end{cases}
\end{equation}
\;\\  
where\\
\;\\
\begin{equation}
\label{r56}
d\varphi _\pm =\frac{d\varphi _\pm}{d\tilde f}{d\tilde f} = \Big ( \mp \frac{1}{2\nu } \frac{1}{\sqrt{1 - \tilde {f} ^2}} + \frac{\alpha }{2\nu } 
\frac{1}{\sqrt{1 - (\alpha \tilde f +\beta )^2}} \Big ) d\tilde f
\end{equation}
\;\\
The part of the integral (\ref{r53}) corresponding to the second term (on the RHS) of the differential (\ref{r56}) (responsible for non elementary character of the integral) vanishes on the interval $(\varphi _{min}, \varphi _{max})$ and one is left with elementary integral (corresponding to the first term in the differential). The angular actions computed in this way read
\;\\
\begin{equation}
\label{r57}
J_\varphi (L) = \frac{\sqrt{2}}{2} \frac{m}{n} \Big( \sqrt{L+\gamma } - \sqrt{c_0 + \gamma }\Big )
\end{equation}
\;\\ 
for the angular potentials $c(\varphi )$ corresponding to the radial $a^k(r)$ ones of oscillator types and
\;\\
\begin{equation}
\label{r58}
J_\varphi (L) = \frac{\sqrt{2}}{2} \frac{m}{n}\Big ( \sqrt{L+ B + \sqrt{F} }  + \sqrt{L+ B - \sqrt{F} }- \sqrt{c_0 + B + \sqrt{F}} - \sqrt{c_0 + B - \sqrt{F}} \Big )
\end{equation}
\,\\
for the $c(\varphi )$ corresponding to the radial $a^k(r)$ of generalized Kepler types.\\
It follows from eqs. (\ref{r51}), (\ref{r57}), (\ref{r52}) and (\ref{r58}) that, as it should be, the linear combinations of actions 
\[
mJ_r^k(E,L) + n J_\varphi (L)
\]
are functions of energy only.
\;\\
\;\\

\section{Superconstants of motion}

Finally, we address a question of the additional (super)constant of motion. Its very existence is built into the framework we are working in. This is due to the well known fact that for any linear combination of action variables with integer coefficients entering the Hamiltonian there exists a globally defined conserved quantity. In particular, the constant of motion corresponding to the linear combination $mJ_r^k +nJ_\varphi $ $m,n \in  Z$
of radial and angular actions entering the Hamiltonian of our system can be written as a real or imaginary part of the quantity 
\;\\
\begin{equation}
\label{r59}
C^k = h(J_r^k,J_\varphi ) e ^{i(m \Psi _\varphi - n \Psi _r^k)}
\end{equation}
\;\\
Here, $h$ is an arbitrary, smooth function of actions $J_r^k$, $J_\varphi$ while $\Psi _\varphi$, $\Psi _r^k$ denote conjugated angle variables, respectively. It readily follows from equations of motion that $C^k$ is conserved.\\
Although, an explicit formula for the extra constant of motion is not necessary to prove the superintegrability, it (if available) not only completes a description but also may provide important and useful information concerning a Poisson algebra of first integrals as well as their dependence on momenta.\\
To obtain the relevant explicit expressions we use eqs.(\ref{r8}) and  (\ref{r9}) to write the (non-single valued) conserved phase $ \phi ^k \equiv m \Psi _\varphi - n \Psi _r^k$ as 
\;\\
\;\\
\begin{equation}
\label{r60}
\phi ^k = \frac{\sqrt{2}}{2}m\frac{dL}{dJ_\varphi }\big (Z(\varphi ) - Y^k(r) \big )
\end{equation}
\;\\
\;\\
where, $Z(\varphi) $, $Y(r)$ denote the following integrals\\
\;\\
\;\\
\begin{equation}
\label{r61}
\begin{split}
& Z(\varphi ) = \int_{\varphi _{min}}^{\varphi }\frac{d\varphi }{\sqrt{L - c(\varphi )}}\\
\;\\
&Y^k(r) = \int_{r_{min}}^{r}
\frac{dr}{s_k^{2}(r)\sqrt{E- a^{k}(r) - \frac{L}{s_k^{2}(r)}} } 
\end{split}
\end{equation}
\;\\
\;\\
while $L(J_\varphi )$ is the inverse map to  $J_\varphi = J_\varphi(L)$  given by eq.(\ref{r57}) or (\ref{r58}), respectively.\\
It appears that the $Z(\varphi )$ integrals unlike $Y^k(r)$ ones are non-elementary, in general. However, performing substitution (\ref{r54}) results in elliptic integrals for the general angular potential given by eq.(\ref{r37}) and for the potential $c(\varphi )$ given by eq.(\ref{r50}). The first corresponds to the radial potential of oscillator/Kepler type while the second one to the radial potential of generalized Kepler type. \\
\;\\
The final explicit formulas for the preserved phase $\Phi ^k$ are very complicated and need further analysis concerning their structure as well as possibility
of writing them in a simpler form.\\
For the models with the radial oscillator potential and the angular $c(\varphi) $ one given by eq.(\ref{r37}) the $\Phi$ constant reads\\
\;\\
\begin{equation}
\label{r62}
\begin{split}
& \Phi = m \Big ( \frac{1}{2}\arcsin{\frac{(L + \gamma )f(\varphi ) - (\gamma + c_0)}{L - c_0}} \\
\;\\
& + 2 \alpha \frac{c_0 + \gamma }{\sqrt{(L + \gamma )(1 + (\alpha  - \beta )}}\quad\frac{\Pi(\Lambda,  \Omega,  \Upsilon)}{\sqrt{(1 - (\alpha  - \beta ))(L + \gamma ) + 2\alpha (c_0 + \gamma )}} \Big )\\
\;\\
&- n \arcsin{\frac{Er^2 -  2(L + \gamma )}{\sqrt{E - 4\omega (L + \gamma )}r^2}} \\
\end{split}
\end{equation}
\;\\
\;\\
where
\;\\
\;\\
\begin{equation}
\label{r63}
\begin{split}
& \Lambda = \arcsin{\sqrt{\frac{(1 + (\alpha  - \beta ))}{(1 - (\alpha  - \beta ))(L + \gamma ) - 2\alpha (\gamma  + c_0)} \quad
\frac{(1 + f(\varphi ))(L + \gamma ) -2(\gamma  +  c_0)}{(1 + f(\varphi ))}}}\\
\;\\
&\Omega = \frac{(1 + (\alpha  - \beta ))(L + \gamma ) - 2\alpha (\gamma  + c_0)}{(1 - (\alpha  - \beta ))(L + \gamma )}\\
\;\\
&\Upsilon  = \sqrt{\frac{(1 + (\alpha  - \beta ))(L + \gamma ) - 2\alpha (\gamma  + c_0)}{(1 - (\alpha  - \beta ))(L + \gamma ) - 2\alpha (\gamma  + c_0)}\quad \frac{1 - (\alpha  - \beta )}{1 + (\alpha  - \beta )}}\\
\end{split}
\end{equation}
\;\\
and $\Pi(\Lambda,  \Omega,  \Upsilon)$ denote the elliptic integral of third kind defined by the formula \cite{l41aa}\\
\;\\
\begin{equation}
\label{r64}
\Pi(\Lambda,  \Omega,  \Upsilon) = \int^\Lambda  _0\frac{dx}{(1 - \Omega \sin^2{x})\sqrt{(1 - \Upsilon ^2 \sin^2{x})}}
\end{equation}
\;\\
$f(\varphi )$ is given by eq.(\ref{r35}).\\
In the case of the Kepler radial potential and the angular potential given by eq.(\ref{r50}) a formula for the the phase $\Phi $
is even more complicated. We present it in the Appendix.\\

\section{Summary}

We have studied the superintegrability  (i.e. the existence of more constant of motion then degrees of freedom) of Hamiltonian systems defined on two-dimensional configuration space of constant curvature and separating in (geodesic) polar coordinates.\\
Classically, as derived in Refs.(\cite{l6}) and (\cite{l29}) the fairly (most?) general class of such systems is defined by two types of allowed radial potentials (oscillator and (generalized) Kepler ones, see eqs. (\ref{r15})  and (\ref{r16})) and by corresponding families of isoperiodic angular potentials $c(\varphi)$. The period of angular motion in the given family is determined by the type of the corresponding radial potential. However, the angular potential itself is given implicitly (up to a function) by, in general, transcendental equation (\ref{r20}). We have been interested in deriving the explicit formulas for the angular potentials providing superintegrable dynamics.\\
Such depending on two  $\alpha $, $\beta $ parameters potentials given in terms of elementary functions are generated by inserting eq.(\ref{r35}) into eqs.(\ref{r26}) or (\ref{r27}), respectively. The allowed $\alpha $, $\beta $ parameters are constrained by the condition $ |\alpha | + |\beta | \leq 1$\\
The resulting general formulas for potentials $c(\varphi)$ (see eqs. (\ref{r37})  and (\ref{r50})) are complicated but in the cases corresponding to the radial potentials of oscillator/Kepler type they significantly simplify for  $\alpha $, $\beta $ living on the boundary $ |\alpha | + |\beta | = 1$. Then one gets the famous Poschl-Teller potential  (see eqs. (\ref{r39}), (\ref{r40})  and (\ref{r43})) or potentials described by eqs. (\ref{r45}) or (\ref{r47}), respectively. Therefore, the general angular potentials given by  eq. (\ref{r37}) can be considered as 2-parameters generalization of Poschl-Teller potential.\\
As mentioned above combining angular Poschl-Teller potential with the radial one of oscillator type (given by  eq. (\ref{r15}))results in the TTW model on the plane (for $ k = 0 $) or its curved counterparts (for $k \neq 0 $). Taking instead of the radial oscillator type potential the Kepler type one (given by  eq. (\ref{r16}) with $F = 0)$) provides PW model on the plane (for $ k = 0 $) or curved spaces (for $k \neq 0 $). In this way, the Hamiltonians with the radial potentials of  oscillator or Kepler type and the general angular potential (\ref{r37}) can be thought as the generalization of TTW and PW Hamiltonians, respectively, on Euclidean plane (for $ k = 0 $) or the curved spaces of constant curvature (for $k \neq 0 $). \\
Another relatively simple forms of angular potentials for both types of the radial ones correspond to the choice $\beta = 0 $. The relevant expressions for potentials are given by eqs. (\ref{r49})  and (\ref{r50}).\\
The superintegability of the considered systems was verified in the independent way by computing the relevant angular and radial action variables and checking that corresponding Hamiltonians depend on a linear combination of the actions with integer coefficients.\\
 The description of  the generalized TTW and PW models have been completed by computing explicitly the additional constant of motion (given by eq. (\ref{r59}), (\ref{r62}), (\ref{r63}) for $k = 0$). It still requires some analysis but it seems rather unlikely that the models are, in general, polynomially superintegrable.\\
The explicit superconstant of motion was also computed for the model with generalized radial Kepler potential on the plane and angular potential given by eq. (\ref{r50}).\\
We conclude with some remarks concerning the quantum mechanical counterparts of the classically superintegrble systems considered above. The classical superintegrability is not broken in the semiclassical WKB approximation $0(\hbar )$ and what is more there exist several nontrivial models (for instance, various versions of Calogero-Moser systems  \cite{l41a}, \cite{l41b},\cite{l41c}, \cite{l41d}, \cite{l41e}, \cite{l41f}, TTW or PW models) which are superintegrable both classically and quantum mechanically. However, in general, higher order corrections in $\hbar $ result in "quantum anomalies" breaking symmetry and spectrum degeneracy of $0(\hbar )$ superintegrable models. It might be interesting to analyze this phenomena in the spirit of Refs.\cite{l42} and \cite{l45} using WKB method valid for any analytical potential \cite{l43}, \cite{l44}.\\
\;\\
\;\\

{\bf Acknowledgments}
The authors would like to thank Piotr Kosi\'nski for helpful discussions and reading the manuscript. Useful remarks of Krzysztof Andrzejewski, Pawel Ma\'slanka and Bartek Zieli\'nski are kindly acknowledged.\\
\;\\
\;\\

\section{Appendix }
  
The formula for the the phase $\Phi $ for the Kepler radial potential and the angular potential given by eq.(\ref{r50})
reads

\begin{equation}
\label{r65}
\phi  = 2n \frac{\sqrt{(L + B)^2 - F}}{\sqrt{L + B - \sqrt{F}} + \sqrt{L + B + \sqrt{F}}}\big ( Z(\varphi ) - Y(r)\big )
\end{equation}
\;\\
where
\begin{equation}
\label{r66}
\begin{split}
& Y(r) = \frac{1}{2\sqrt{L + B - \sqrt{F}}}\arcsin{\Big ( \frac{1}{\sqrt{\Delta }}\big (1 + 2\frac{\sqrt{F}}{D}E \big ) - \frac{2(L + B - \sqrt{F})}{\sqrt{Dr^2 + F} - \sqrt{F}}}\Big ) +\\
\;\\
& \frac{1}{2\sqrt{L + B + \sqrt{F}}}\arcsin{\Big ( \frac{1}{\sqrt{\Delta }}\big (1 - 2\frac{\sqrt{F}}{D}E \big ) - \frac{2(L + B + \sqrt{F})}{\sqrt{Dr^2 + F} + \sqrt{F}}\Big )} 
\end{split}
\end{equation}
\;\\
with\\
\;\\
 $\Delta  = \displaystyle{1 + 4\frac{E}{D}\big ( B + L + \frac{EF}{D}\big )}$;\\
\;\\
\begin{equation}
\label{r67}
\begin{split}
& Z(\varphi ) = \frac{m}{n}\bigg ( \frac{1}{4\sqrt{L + B + \sqrt{F}}}\Big (\frac{\pi }{2} - \arcsin{\frac{P_+(\rho + \sqrt{F}) +2Q_+}{(\rho + \sqrt{F})\sqrt{\Delta }}}\Big ) + \\
\;\\
& \frac{1}{4\sqrt{L + B - \sqrt{F}}}\Big ( \frac{\pi }{2} - \arcsin{\frac{P_-(\rho - \sqrt{F}) +2Q_-}{(\rho - \sqrt{F})\sqrt{\Delta }}}\Big ) +\\
\;\\
& \frac{\alpha }{2}\sqrt{\frac{(J + B)^2 - F}{J- L}}\frac{a - b}{\sqrt{(a + b)(b-d)}}\\
&\Big ( \frac{1}{a - \sqrt{F}}\Pi \big (\mu ,\frac{(b - a)(d - \sqrt{F})}{(b - d)(a - \sqrt{F})}, \zeta \big ) + \frac{1}{a + \sqrt{F}}\Pi \big (\mu ,\frac{(b - a)(d + \sqrt{F})}{(b - d)(a + \sqrt{F})}, \zeta \big )\Big )\bigg )
\end{split}
\end{equation}
\;\\
with\\
\;\\
$a = \displaystyle{\frac{(J + B)(L+ B) -F }{L- J}}$\\
\;\\
$b = \sqrt{F + \alpha ^2((J + B)^2 - F)}$\\
\;\\
$d = - (J + B)$\\
\;\\
$\rho (\varphi ) = \displaystyle{\frac{1}{f(\varphi )}\sqrt{(J + B)^2 - F(1 + f^2(\varphi ))}}$\\
\;\\
$\mu  = \displaystyle{\arcsin{\sqrt{\frac{(b - d)(a -\rho (\varphi ) )}{(b - a)(a -\rho (\varphi ))}}}}$\\
\;\\
$\zeta  = \displaystyle{\sqrt{\frac{(b - a)(b + d)}{(b + a)(b - d)}}}$\\
\;\\
$P_{\pm} = (J+B)^2 -F \pm 2(J-L)\sqrt F$\\
\;\\
$Q_{\pm} = \left ( F-(J+B)^2\right ) \left ( L+B \pm \sqrt F\right )$.
 



\begin{thebibliography}{99}






\bibitem{l1}Nekhoroshev N. N. 1972 Action-angle variables and their generalization Trans. Moscow.Math. Soc. 26 180-198 2

\bibitem{l2}Arnold  V. I., 1978 Mathematical Methods of Classical Mechanics (New York: Springer-
Verlag)\\
Perelomov A. M. ( 1990) Integrable Systems of Classical Mechanics and Lie Algebras, Birkhauser,\\
Babelon O., Bernard D., Talon M. (2007), Introduction to Classical Integrable Systems (Cambridge Monographs on Mathematical Physics)

\bibitem{l3}Miller W. Jr, Post S and Winternitz P 2013 Classical and quantum superintegrability with applications J. Phys. A-Math. Gen. 46 423001\\
 Kalnins E. G., Kress J. M. and Miller W. Jr. 2018 Separation of variables and Superintetgrability: The symmetry of solvable systems (UK, ISBN: 978-0-7503-1314-8: Instititute of Physics)




\bibitem{l4}P. Winternitz, J. Smorodinsky, M. Uhlir and I. Fris, Symmetry groups in classical and quantum mechanics.Yad. Fiz, 4 625-635, 1966 ( English translation Sov. J. Nucl. Phys. 4, 444-450 (1967))


\bibitem{l5}I. Fris, V. Mandrosov, J. Smorodinsky, M. Uhlir, and P. Winternitz, On higher symmetries in quantum mechanics. Phys. Lett., 16:354–356 (1965)

\bibitem{l6}Onofri E. and Pauri M., 1978 Search for periodic Hamiltonian flows: A generalized Bertrand's
theorem J. Math. Phys. 19 1850


\bibitem{l7}Higgs P. W. (1979) Dynamical symmetries in a spherical geometry I J.Phys. A: Math. Gen. 12, 309

\bibitem{l8}Leemon H. I.( 1979) Dynamical symmetries in a spherical geometry II J. Phys. A: Math. Gen. 12, 489


 \bibitem{l9}Granovskii Y. I., Zhedanov A. S., Lutsenko I. M., (1992) Quadratic algebras and dynamics in curved spaces. ii. The Kepler problem Theoret.and Math. Phys. 91, 604

\bibitem{l10}D. Bonatsos, C. Daskaloyannis and K. Kokkotas, Quantum algebraic desription of quantum superintegrable systems in 2 dimensions. Phys. Rev. A 48(5), R23407-R3410 (1993)

\bibitem{l11}Ranada M. F . , Santander M.,(1999) Superintegrable systems on the two-dimensional sphere S2 and the hyperbolic plane H2 J. Math. Phys.40, 5026


\bibitem{l12}C. Daskaloyannis, Quadratic Poisson algebras of two-dimensional classical superintegrable systems and quadratic algebras of quantum superintegrable systems, J. Math. Phys. 42 1100–1119 (2001)

\bibitem{l13}Saksida P. (2001) Integrable anharmonic oscillators on spheres and hyperbolic spaces Nonlinearity 14, 977

\bibitem{l14}Ranada M. F. , Santander M. ( 2002) On harmonic oscillators on the two-dimensional sphere S2 and the hyperbolic plane H2 J. Math. Phys. 43, 431

\bibitem{l15}Gravel S. and Winternitz P., 2002 Superintegrability with third-order integrals in quantum
and classical mechanics J. Math. Phys. 43 5902–12

\bibitem{l16}Ranada M. F. and Santander M., (2003) On harmonic oscillators on the two-dimensional sphere S2 and the hyperbolic plane H2 II J. Math. Phys. 44, 2149

\bibitem{l17}Borisov A. V. and Mamaev I. S., Superintegrable Systems on a Sphere, Regul. Chaotic Dyn., 2005, vol. 10, no. 3, pp. 257–266.

\bibitem{l18}Kalnins E G, Kress J M and Miller W Jr 2005 Second-order superintegrable systems in conformally flat spaces. I. Two-dimensional classical structure theory J. Math. Phys. 46 053509

\bibitem{l19}Shchepetilov A.V.(2005) Comment on Central potentials on spaces of constant curvature: The Kepler problem on the two-dimensional sphere S2 and the hyperbolic plane H2 [J.Math. Phys. 46, 052702], J. Math.Phys. 46, 114101

\bibitem{l20}Borisov A. V., Kilin A.A., and Mamaev I. S., Superintegrable System on a Sphere with the Integral of Higher Degree, Regul. Chaotic Dyn., 2009, vol. 14, no. 6, pp. 615–620.

\bibitem{l21}F. Tremblay, A. V. Turbiner, and P. Winternitz, An infinite family of solvable and integrable quantum systems on a plane. J.Phys.A.Math.Theor., 42(24):242001, 2009.

 \bibitem{l22}F. Tremblay, A. V. Turbiner, and P. Winternitz, Periodic orbits for a family of classical superintegrable systems. J.Phys.A.Math.Theor., 43(1):015202, 2010. 

\bibitem{l22a}Post S.,  Winternitz P. (2010) An infinite family of superintegrable deformations of
the Coulomb potential J. Phys. A: Math. Theor. 42, 222001

\bibitem{l23}Maciejewski A. J, Przybylska M. and Yoshida H. (2010) Necessary conditions for classical super-integrability of a certain family of potentials in constant curvature spaces J.Phys. A 43, 382001

\bibitem{l24}Gonera C. 2012 On the superintegrability of the TTW model Phys. Lett. A 376 2341–3

\bibitem{l25}Hakobyan T., Lechtenfeld O., Nersessian A., Saghatelian A. and Yeghikyan V.,(2012) Integrable generalizations of oscillator and Coulomb systems via action-angle variables,Phys. Lett. A 376, 679

\bibitem{l26}E. Celeghini, S. Kuru, J. Negro and M.A. del Olmo, A unified approach to quantum and classical TTW systems based on factorization Ann. Phys. 332 27-37(2013)

\bibitem{l26a}Galajinsky A.,Lechtenfeld O.,On two-dimensional integrable models with a cubic or quartic integral of motion,
JHEP 1309, (2013),113 


\bibitem{l27}Ranada M F 2013 Higher order superintegrability of separable potentials with a new
approach to the Post-Winternitz system J. Phys. A-Math. Theor. 46, 125206

\bibitem{l28}Ivan A. Bizyaev, Alexey V. Borisov, Ivan S. Mamaev, "Superintegrable Generalizations of the Kepler and Hook Problems", Regul. Chaotic Dyn., 19:3 (2014), 415–434

\bibitem{l28a}Manuel F. Ranada (2014), The Tremblay–Turbiner–Winternitz system on spherical and hyperbolic spaces: superintegrability, curvature-dependent formalism and complex factorization, J. Phys. A: Math. Theor. 47 165203

\bibitem{l29}Gonera C. and Kaszubska M. 2014 Superintegrable systems on spaces of constant curvature
Annals of Physics 346 91-102

\bibitem{l30}Hakobyan T., Nersessian A. and Shmavonyan H., 2017 Lobachevsky geometry in TTW and
PW systems Phys. Atomic Nuclei 80 598–604

\bibitem{l31}P. Iliev, Symmetry algebra for the generic superintegrable system on the sphere. J. High Energy Phys. 2, 44  (2018)

\bibitem{l32}I. Marquette and P. Winternitz. Superintegrable systems with third order integrals of motion. J. Phys. A-Math.Theor., 41(30):303031 (10 pages), 2008.

\bibitem{l33}Tremblay F. and Winternitz P. 2010 Third-order superintegrable systems separating in polar coordinates J. Phys. A-Math. Theor. 43 175206

\bibitem{l34}Post S. and Winternitz P. 2015 General Nth-order integrals of motion in the Euclidean plane J. Phys. A-Math. Theor. 48 405201

\bibitem{l35}A. M. Escobar-Ruiz and J. C. Lopez Vieyra and P. Winternitz, Fourth order superintegrable systems separating in polar coordinates. I. Exotic potentials. J. Phys. A : Math. and Theor. 50(49), 495206, 2017.

\bibitem{l36} A. M. Escobar-Ruiz, J. C. Lopez Vieyra, P. Winternitz and I. Yurdusen, Fourth-order superintegrable systems separating in polar coordinates. II. Standard potentials. J. Phys. A: Math. and Theor. 51(45),455202, 2018.

 \bibitem{l37}A. M. Escobar-Ruiz, P. Winternitz and I. Yurdusen, General N-th order superintegrable systems separating in polar coordinates. J. Phys. A : Math. and Theor. 51(40):40LT01, 2018
 
\bibitem{l38}Landau L.,Lifshitz E., (1976) Mechanics, Pergamon Press,



\bibitem{l39}Tsiganov  A.V. (2019), Superintegrable systems with algebraic and rational integrals of motion, Theoretical and Mathematical Physics 199 (2), 659-674

\bibitem{l40}Tsiganov A.V., Elliptic curve arithmetic and superintegrable systems, accepted to Physica Scripta, arXiv: 1810.11991, 2018.

\bibitem{l41}Tsiganov A.V.,(2019) The Kepler problem: polynomial algebra of non-polynomial first integrals, preprint, arXiv:1903.08846v1 [nlin.SI] 21 Mar 2019

\bibitem{l41aa} I.S. Gradshteyn and I.M. Ryzhik, Table of integrals, series and products, Academic Press, New York and London, 1965
\bibitem{l41a}V. B Kuznetsov, Hidden symmetry of the quantum Calogero-Moser system, Phys. Lett.
A 218 (1996) 212

\bibitem{l41b}C. Gonera,(1998) A note on superintegrability of the quantum Calogero model, Physics Letters A 237 (6), 365-368

\bibitem{l41c}R. Caseiro, J.-P. Francoise, R. Sasaki, Quadratic algebra associated with rational Calogero-Moser models, J. Math. Phys. 42, 5329–5340 (2001)

\bibitem{l41d}N. Reshetikhin, Degenerate integrability of the spin Calogero-Moser systems and the duality with the spin Ruijsenaars systems, Lett. Math. Phys. 63(1), 55–71 (2003).

\bibitem{l41e}N. Reshetikhin, Degenerate integrability of quantum spin Calogero - Moser systems, Lett. Math. Phys. 107(1), 187–200 (2017).



\bibitem{l41f}M. Feigin, T. Hakobyan, Algebra of Dunkl Laplace-Runge-Lenz vector, arXiv:1907.06706, 2019



\bibitem{l42}J. Dorignac, On the quantum spectrum of isochronous potentials, J. Phys; A: Math. Gen.,38, p. 6183-6210 (2005)

\bibitem{l43}Robnik M. and Salasnich L. 1997, WKB to all orders and the accuracy of the semiclassical quantization, J. Phys. A: Math. Gen., 30, 1711.

\bibitem{l44}Robnik M. and Romanovski V.G. 2000, Some properties of WKB series, J. Phys. A: Math. Gen., 33, 5093

\bibitem{l45}A. Raouf Chouikha, "On the Isochronous analytic motions and the quantum spectrum" accepted to Physica Scripta,  arXiv: 1811.08315 [math-ph].
	













 

\end{thebibliography}
\end{document}